\documentclass[final,5p,times,twocolumn]{elsarticle}

\usepackage{amssymb}

\journal{Material Science in Semiconductor Processing}

\begin{document}

\begin{frontmatter}



\title{Optical properties of wurtzite GaN/AlN quantum dots grown on non-polar planes: the effect of stacking faults in the reduction of the internal electric field}


\author{J. A. Budagosky, N. Garro, A. Cros, and A. Garc\'\i a-Crist\'obal}

\address{Institut de Ci\`encia dels Materials (ICMUV), Universitat de
Val\`encia, E-46071 Val\`encia, Spain}

\author{S. Founta and B. Daudin}

\address{CEA, INAC-SP2M, "Nanophysique et semiconducteurs" group, F-38000 Grenoble, France}

\begin{abstract}
The optical emission of non-polar GaN/AlN quantum dots has been investigated. The presence of stacking faults inside these quantum dots is evidenced in the dependence of the photoluminescence with temperature and excitation power. A theoretical model for the electronic structure and optical properties of non-polar quantum dots, taking into account their realistic shapes, is presented which predicts a substantial reduction of the internal electric field but a persisting quantum confined Stark effect, comparable to that of polar GaN/AlN quantum dots. Modeling the effect of a 3 monolayer stacking fault inside the quantum dot, which acts as zinc-blende inclusion into the wurtzite matrix, results in an additional 30\% reduction of the internal electric field and gives a better account of the observed optical features.

\end{abstract}


\end{frontmatter}


\section{Introduction}
The current interest in group-III nitride quantum dots (QDs) is
mostly related to the possibility of developing a new generation
of optoelectronic devices operating in the ultraviolet wavelength
range \cite{nakamura00}. As a matter of fact, self-assembled GaN/AlN QDs
grown in the Stranski-Krastanov mode by molecular beam epitaxy (MBE),
which present hexagonal wurtzite (WZ) structure preferably, are free of
structural defects, such as dislocations \cite{daudin97}. The great potential of these systems is somehow frustrated by the
presence of strong built-in electrostatic fields which induce the
quantum confined Stark effect \cite{miller84} (QCSE), i.e. the
emission exhibits large red-shifts and the oscillator strengths of
the optical transitions are reduced considerably due to electron
and hole spatial separation. These fields are originated by the
difference in the macroscopic polarization between the dot and the
barrier materials and are mainly parallel to the polar $c$-axis of
the WZ structure \cite{bernardini97}. Non-polar heterostructures have been grown with the aim of
reducing or suppressing the internal electric field.
In the case of non-polar quantum wells, where the macroscopic
polarization lays on the plane of the well and undergoes no discontinuity, built-in electric fields
are completely suppressed \cite{Waltereit00}. The lack of electric field effects reported for GaN/AlN QDs
grown on the a-plane by MBE is more intriguing. These QDs present strong emission in the UV with fast lifetimes indicating an effective suppression of the QCSE \cite{founta05,garro05,rol07}.

In this paper we study the emission of a-plane GaN/AlN QDs and compare it with the predictions of a theoretical model which takes into account the realistic arrow-head shapes of these systems \cite{founta07}.
The absorption spectrum is calculated using a plane-wave expansion method and a $8\times 8$ \textbf{k}$\cdot$\textbf{p} model. The comparison of polar and non-polar QD absorption shows that the latter are still dominated by QCSE, in agreement with previous theoretical models \cite{schulz09,marquardt09,schuh12,schulz12}, not fitting, however, our experimental observations. The influence of basal plane stacking faults on the QD optical properties is revealed in the temperature and power dependence of their emission. We show how such stacking faults, acting as zinc-blende (ZB) inclusions, modify the electric polarization inside the QDs and can account for additional electric field reductions. Such reduction, united to the effects of Coulomb correlation \cite{schuh12} or the realistic shape of the dots \cite{founta07,schulz12}, could complete the picture of the QCSE suppression in non-polar QDs.

\section{Experimental results}\label{expdescription}

We have investigated a sample consisting of a 18-period multi-layer of GaN/AlN QDs grown by MBE on commercial $a$-plane 6H-SiC substrate. Self-assembled GaN QDs ($a$-QDs) formed by the Stranski-Krastanow growth method at 700 $^\circ$C under Ga-rich atmosphere. The mean diameter and height of the GaN QDs are $20$ nm and 3 nm, respectively. More details about the growth of this sample can be found elsewhere \cite{garro05}.

\begin{figure}[t]
\centering
\includegraphics[trim = 2cm 2cm 2cm 2cm, clip, scale=0.45]{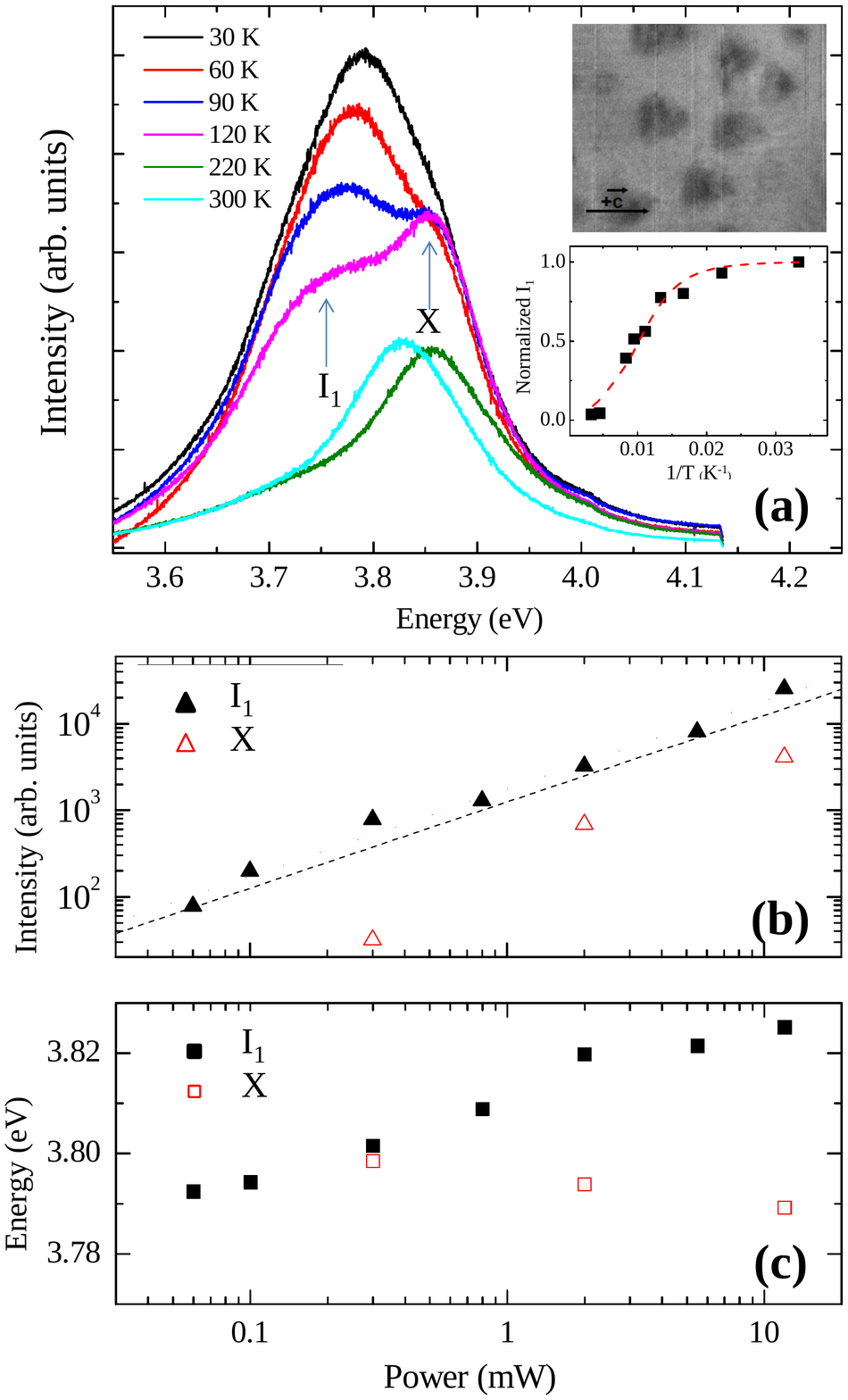}
\caption{(Color on-line) (a) PL spectra of the GaN $a$-QDs as a function of the sample temperature. The upper inset shows a $100\times80$ nm$^2$ plan-view TEM image where BSFs crossing the AlN barriers and most of the $a$-QDs appear as vertical lines, as extracted from Ref. \cite{founta07}. The lower inset contains an Arrhenius plot of the intensity of the $I_1$ peak providing an activation energy of $30\pm3$ meV.
The dependence of the peak intensity (b) and energy (c) on the excitation power of the $I_1$ and $X$ transitions. The dashed line in (b) corresponds to the linear dependence.} \label{fig-1ex}
\end{figure}

The spectral content and the dynamics of the emission of the $a$-QDs have been explored by time-integrated and time-resolved photoluminescence (PL) spectroscopy. Figure \ref{fig-1ex} (a) shows the $a$-QD PL spectra recorded between 30 and 300 K. We observe that the emission peaks above the GaN band-gap energy indicating an effective suppression of the QCSE, in agreement with previous reports \cite{founta05,garro05,rol07}. On the other hand, two transitions can be identified (labeled as $I_1$ and $X$) peaking at 3.785 eV and 3.87 eV at 30 K, which evolve differently with temperature. While $I_1$ dominates the emission at low temperatures, its intensity decays rapidly as the temperature is raised, as shown by the Arrhenius plot insetted in Fig. \ref{fig-1ex} (a). At room temperature, just one gaussian peak corresponding to transition $X$ is observed. The dynamics of the PL (not shown) is also different for $X$ and $I_1$ with decay times of 1.23 and 2.38 ns, respectively. In order to study the dependence of the two transitions on the excitation power, experiments have been carried out at low and room temperature, when the PL is dominated by $I_1$ and $X$, respectively. The power dependence of the intensity and the PL peak energies are exemplified in Figs. \ref{fig-1ex} (b) and (c). The integrated intensities increase with power almost linearly in both cases. The peak energy, however, blue-shifts clearly with increasing power for the $I_1$ transition, while being nearly constant for $X$.
The overall phenomenology reported for transition $I_1$ is compatible with that attributed to basal stacking faults (BSFs) \cite{corfdir09,jacopin11}. Non-polar QDs can be often crossed by BSFs which are common in $a$-plane AlN, as shown in TEM image of the upper inset of Fig. \ref{fig-1ex} (a).

BSFs are ZB inclusions in the WZ matrix and can be regarded as type-II quantum wells for electrons. Holes are in the surrounding WZ area and stay bound to the electrons thanks to Coulomb attraction. Therefore, transition $I_1$ in Fig. \ref{fig-1ex} corresponds to the radiative recombination of such type-II excitons. Thermal excitation can delocalize the holes by dissociating the exciton. Actually, the activation energy obtained experimentally, $30 \pm 3$ meV, matches the binding energy of the exciton \cite{corfdir09}. Increasing the excitation power, on the other hand, leads to the band-filling of the electron subbands and the partial screening of the electric field in the inclusion, which is originated by the discontinuity of the polarization at the WZ/ZB interface \cite{jacopin11}. This can explain the blue-shift of the $I_1$ PL peak shown in Fig. \ref{fig-1ex} (c).

\section{Theoretical results}\label{modeldescription}

\subsection{Description of the model}

We have implemented a model for the optical properties of GaN/AlN QDs with arbitrary geometry and both polar and non-polar orientations. The first step in our model is the calculation of the three-dimensional strain and built-in potential distributions in the dot and the surrounding barrier. Our approach to the strain distribution is based in a reformulation of Eshelby's inclusions model \cite{eshelby57}. The displacement, ${\bf u}({\bf r})$, associated to the relaxation of the QD inclusion within the infinite matrix is calculated from
the elastic equilibrium equation,
\begin{equation}\label{equilibriumeq}
\frac{\partial}{\partial x_j}\left[C_{ijkl}({\bf r})\frac{\partial}{\partial x_l}u_k({\bf r})\right]=-\frac{\partial}{\partial x_j}\left[\sigma_{ij}^{(0)}\chi({\bf r})\right]~~,
\end{equation}
where $\chi({\bf r})$ is the characteristic shape function of the inclusion,  $C_{ijkl}({\bf r})$ are the elastic constants of the inhomogeneous system, and $\sigma_{ij}^{(0)}$ the {\it initial stress}, related to the mismatch between the lattice parameters of the matrix and the inclusion. The analysis is simplified by taking the same elastic constant values in the dot and matrix, which is a good approximation for GaN and AlN. Since the shape
of the non-polar QDs can be crucial for their optical properties \cite{schulz12}, we turn to the Fourier space where it is always possible to obtain the analytical solutions for $\tilde{u}_i({\bf q})$ for any QD shape. Finally, the built-in electrostatic potential is computed considering the spontaneous, $\mathbf{P}^{\rm{sp}}=(0,0,P^{\rm{sp}}_z)$, and piezoelectric, $\mathbf{P}^{\rm{pz}}=(P^{\rm{pz}}_x,P^{\rm{pz}}_y,P^{\rm{pz}}_z)$, contributions for
the total material polarization. The latter are calculated assuming homogeneous piezoelectric constants for the inclusion and the matrix. Then, we use these polarizations and the Poisson equation to compute the built-in electrostatic potential,
\begin{equation}\label{potential}
\tilde{\phi}(\mathbf{q})=-\frac{i}{q^2\epsilon_0\epsilon_r}\mathbf{q}\mathbf{\cdot}\tilde{\mathbf{P}}^{\rm{tot}}(\mathbf{q}) \quad .
\end{equation}

The electronic structure is calculated within the multi-band envelope function approximation based on the $8\times 8$ \textbf{k}$\cdot$\textbf{p} Hamiltonian derived by Chuang {\it et al} \cite{chuang96} taking into account the contribution of the strain field and the built-in electrostatic potential. A plane-wave expansion within the first Brillouin zone is performed \cite{yanvoon09}. Excitonic correlation has been taken into account perturbatively.
Then, for each electron-hole pair the corresponding exciton energy is given by $E_{X,ij}=(E^c_j-E^v_i)-E_{b,ij}$, where $E_{b,ij}$ is calculated from the direct matrix element of the electron-hole Coulomb interaction
\begin{equation}\label{eq-22}
E_{b,ij}=-\frac{e^2}{4\pi\varepsilon_r}\int_{\mathbf{r}^e}
\int_{\mathbf{r}^h}\frac{|\psi_j^c(\mathbf{r}^e)|^2|\psi_i^v(\mathbf{r}^h)|^2}{\sqrt{(x_e-x_h)^2+(y_e-y_h)^2+(z_e-z_h)^2}}~~\quad .
\end{equation}
Here, $\psi_j^c$ ($E^c_j$) and $\psi_i^v$ ($E^v_i$) are the wave functions (energies) of the $j^{\rm{th}}$ electron state and the $i^{\rm{th}}$ hole state, respectively. We finally obtain the absorption spectrum from the exciton energies and the computed polarization dependent dipolar matrix elements
\begin{equation}\label{eq-24}
\alpha(\hbar\omega) = \frac{C}{\hbar\omega}\sum_{i,j}f_{T,ij}\frac{\Gamma}{\left(E_{X,ij}-\hbar\omega\right)^2+\Gamma^2}~~~\quad ,
\end{equation}
where $C$ is a constant, $\hbar\omega$ the photon energy, $\Gamma$ provides a phenomenological description of the line broadening, and $f_T$ represents the transition oscillator strength, defined as
\begin{equation}
\label{oscstrength}
f_T=\frac{2\hbar^2}{m_0E_X}\left| \langle\psi_j^c|\mathbf{\hat{e}\cdot p}|\psi_i^v\rangle \right|^2\quad .
\end{equation}
Finally, an important property of the confined electron-hole transitions is their radiative lifetime, $\tau_r$ \cite{rol07}
\begin{equation}\label{lifetime}
\tau_r=\frac{3m_0c^32\pi\hbar^2\epsilon_0}{ne^2E_X^2f_T}\quad ,
\end{equation}
where $n$ is the refraction index of the material. The parameters used in our calculations are those of Ref \cite{andreev00}.

\subsection{Comparative study of the optical properties of polar and non-polar GaN/AlN QDs}

\begin{figure}[t]
\centering
\includegraphics[trim = 2.2cm 1cm -1cm 1cm, clip, scale=0.35]{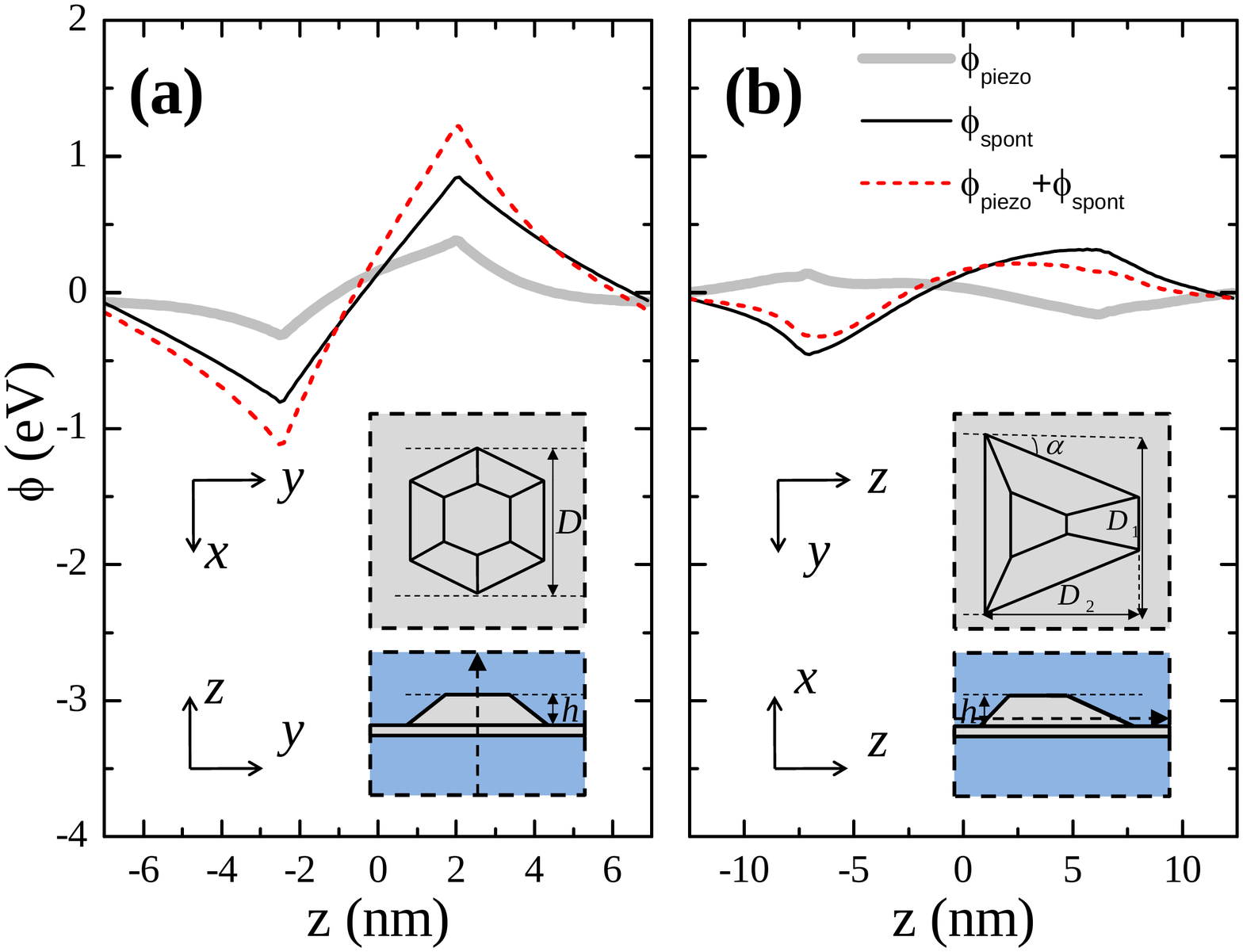}
\caption{(Color on-line) Variation of the electrostatic potential components along the wurtzite $c$-axis of a (a) $c$-QD and a (b) $a$-QD.
The shapes of the QDs for each orientation are sketched in the insets. The dimensions of the $c$-QD are $D = 20$ nm and $h= 4$ nm; and of the $a$-QD, $D_1 = 24$ nm, $D_2 = 16$ nm, $h=3$ nm, and $\alpha= 30^\circ$. The dashed arrows indicate the lines where the potential profiles are taken from.} \label{fig-2t}
\end{figure}

Two different QD orientations are studied, which involve two distinctive QD geometries. Polar QDs, with the dot axis parallel to the $c$-axis ($c$-QDs), are modeled as truncated hexagonal
based pyramids \cite{daudin97}, as depicted in Fig \ref{fig-2t} (a).  On the other hand, $a$-QDs have the truncated trapezoidal based pyramidal shape \cite{founta07} shown in Fig \ref{fig-2t} (b). We shall take the $z$-axis always parallel to the crystal [$0001$] direction. Figure \ref{fig-2t} shows the variation of the electrostatic potential along the $z$-direction for both types of QDs. The contributions from the piezoelectric and the spontaneous polarizations are explicitly displayed. For the $c$-QD, the piezoelectric and spontaneous contributions are found to be of similar magnitude and, more importantly, of the same sign, resulting in a total potential that changes almost $2.4$V in a length of only $4$nm. Notice that, from the electrostatic point of view, a QD of this geometry resembles a planar capacitor, with its oppositely charged sheets separated a distance equal to the QD height. Therefore, the potential difference between the top and bottom of the QD increases almost linearly with height. A similar analysis for the $a$-QD, see Fig. \ref{fig-2t}(b), conveys very different results. First, we notice that the contribution to the potential arising from the spontaneous polarization has almost the same magnitude than that for the $c$-QD. However, with the image of the planar capacitor in mind, we would expect a much larger value. The
value obtained for the potential can be partially understood through geometrical arguments: the GaN-AlN interface along $z$ in the $a$-QD is not perpendicular to the $c$-axis, leading to a reduction in the polarization charge. Additionally, the magnitude of the potential does not increase linearly with distance, since the $a$-QD geometry hardly resembles a planar capacitor. A second striking difference between Figs. \ref{fig-2t}(a) and (b) arises from the piezoelectric contribution to the potential induced by strain which has now opposite sign due to the differences in the strain distribution \cite{garro05}. As a consequence, the total potential is considerably reduced.

\begin{figure}[t]
\centering
\includegraphics[trim = 1cm 0cm 2cm 2cm, clip, scale=0.3]{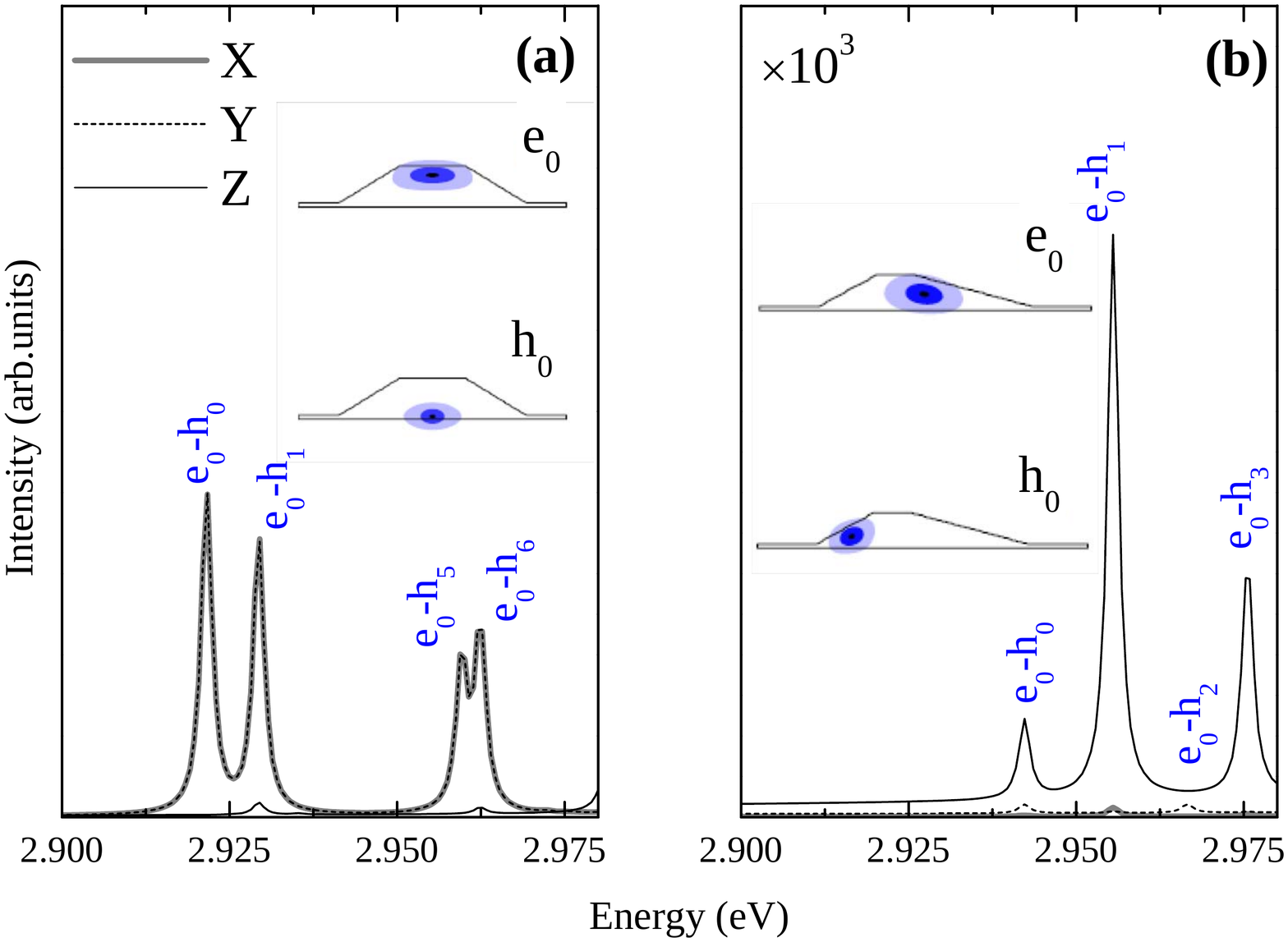}
\caption{(Color on-line) First transitions of the optical absorption spectra of (a) the $c$-QD and (b) $a$-QD as a function of the light polarization.
The labeling for each transitions is realized by the use of the electron and hole levels. The insets show the wavefunctions of the electron and hole ground states for the two geometries.} \label{fig-4t}
\end{figure}

The absorption spectra of  polar and non-polar QDs are compared in Fig \ref{fig-4t}. The labels indicate the electron and hole states involved in the transitions for both orientations. For the $c$-QD, the absorption peaks corresponding to the ground state are well below the GaN band-gap energy ($2.922$ eV and $2.930$ eV). The small energy difference between both peaks is attributed to the band coupling included in the model.
Remarkably, the optical transitions of the $a$-QD are just 20 meV higher than those of the $c$-QD, which can be explained by the smaller volume of the
$a$-QD with respect to the polar case and the consequent increase of the quantum confinement energy. Despite the similarities in terms of energy range, we find profound differences in the two spectra. First of all, the absorption spectrum of $a$-QDs is three orders of magnitude weaker than that of the $c$-QDs. Such a decrease in the oscillator strength is due to the larger spatial separation between the electron and hole states in non-polar QDs, as can be observed in the insets of Fig. \ref{fig-4t} for the ground states of electrons and holes confined in such QDs.
In relation with that, the calculated lifetime of the radiative recombination of the ground transition is $18$ ms for $a$-QDs. This value is four orders of magnitude greater than the lifetime of the ground transition in $c$-QDs, ($2.3\, \mu$s). From the theoretical simulation we can conclude that the QCSE ends up being higher in $a$-QDs than in $c$-QDs.

\section{Discussion}
\label{}

\begin{figure}[t]
\centering
\includegraphics[trim = 2cm 2cm 2cm 2.5cm, clip, scale=0.37]{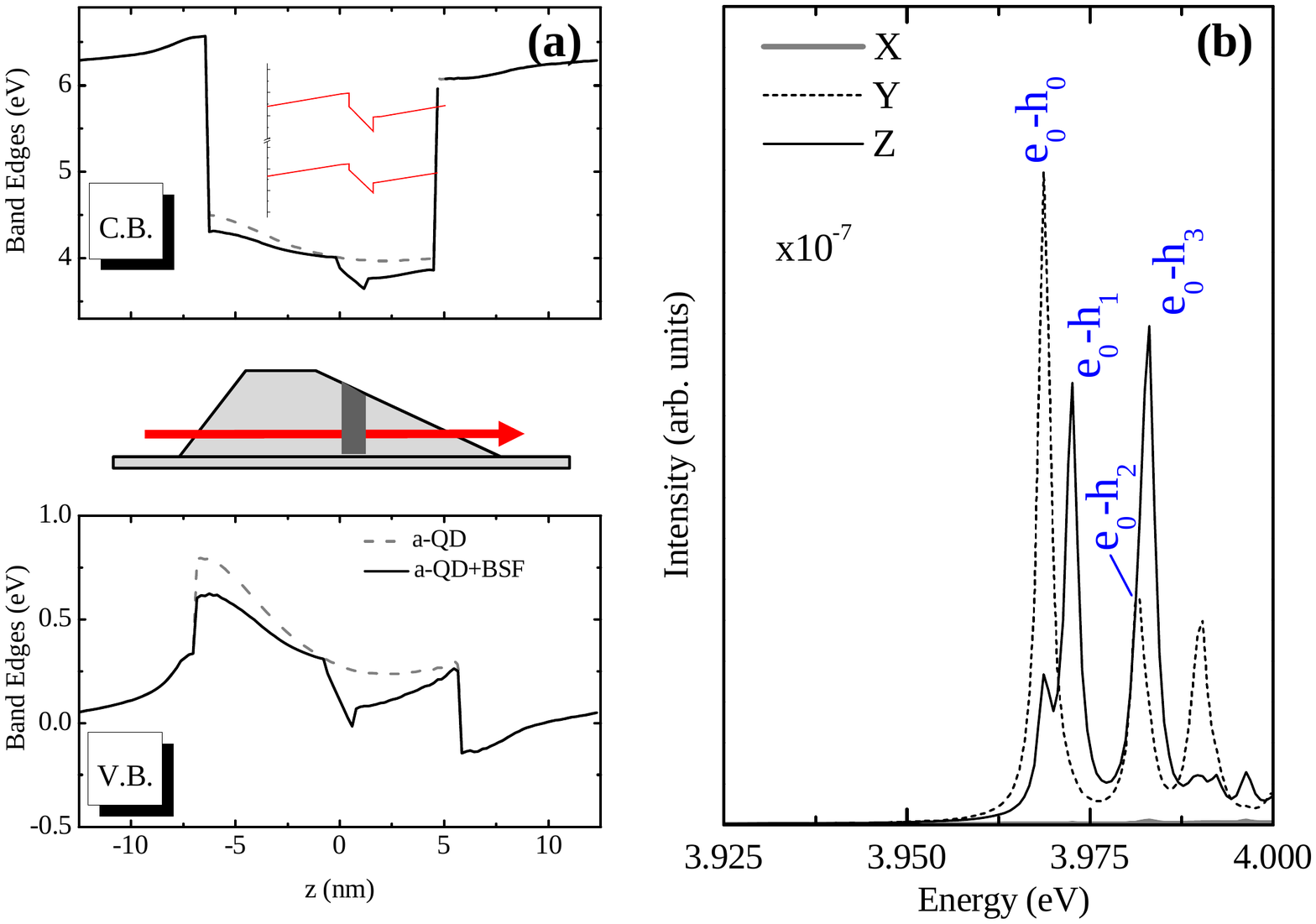}
\caption{(Color on-line) (a) Conduction and valence band-edge profiles for the $a$-QD corresponding to the first states for electrons and holes containing a 3 mono-layer ZB inclusion in the $z=0$ plane. (b) First transitions of the optical absorption spectrum of electric field-free $a$-QD. The scale of the intensity is the same as in Fig. \ref{fig-4t}.} \label{fig-5t}
\end{figure}

The almost full suppression of the QCSE evidenced experimentally cannot be explained by our theoretical simulations. This puts forward the need for other mechanisms that should account for additional reductions of the internal electric fields in non-polar QDs. While some authors have demonstrated that the shape of the dots \cite{schulz12} or the presence of free carriers \cite{hille14} can increase the oscillator strength of the transitions or screen built-in electric fields, no attention has been paid to the existence of structural defects, such as BSFs, revealed in the PL spectra of Fig. \ref{fig-1ex}. The effect of BSFs is twofold: first of all, the band offsets between the WZ matrix and the ZB inclusion generates a thin type-II quantum well ($\Delta E_C =122$ meV and $\Delta E_V=-62$ meV) where only electrons should be confined \cite{corfdir09}; secondly, the lack of spontaneous polarization of the ZB phase (piezoelectric polarizations are comparable in both phases) yields an additional contribution to the electric field. Thus the band-edge profiles of BSFs are those of a triangular quantum well, as sketched in the inset of Fig. \ref{fig-5t} (a) for a 3 mono-layer ZB inclusion, reproducing the calculations of Ref. \cite{jacopin11}. If a similar inclusion was located in the center of an $a$-QD, the band-edge profiles of the QD would be modified by the BSF potential. Figure \ref{fig-5t} (a) shows the comparison of the $a$-QD confinement potential with (solid line) and without (dashed line) the BSF. The presence of the ZB inclusion reduces the internal electric field in the WZ area by around 30\%.
Even more substantial reductions of the built-in electric field are expected for thicker ZB inclusions \cite{jacopin11}. Therefore we can justify the suppression of the QCSE by a combination of effects, including the presence of BSFs. The calculated absorption spectrum of an $a$-QD with zero internal electric field is displayed in Fig. \ref{fig-5t} (b). The energy of the transitions now matches that of the PL peak and, more importantly, the oscillator strength is seven orders of magnitude larger than those obtained for $c$-QDs.

In conclusion, the optical properties of GaN/AlN non-polar quantum dots show a drastic suppression of the quantum confined Stark effect in these systems. Theoretical calculations predict a strong reduction of the internal electric field which does not ensure, however, an effective suppression of the quantum confined Stark effect due to the orientation of non-polar quantum dots. The computation of the potential profile of basal stacking faults inside the quantum dots provides additional reduction of the internal electric field. Excitonic transitions associated to stacking faults can dominate the emission spectrum at low temperatures and moderate excitation powers.




\begin{thebibliography}{00}



\bibitem{nakamura00}
S. Nakamura, S. Pearton, and G. Fasol, {\it The
blue laser diode} (Springer, Berlin Heidelberg, 2000).
\bibitem{daudin97} B. Daudin, F. Widmann, G. Feuillet, Y. Samson, M.
Arlery, and J. L. Rouvi\`ere, Phys. Rev. B {\bf 56}, R7069 (1997).
\bibitem{miller84}  D. A. B. Miller, D. S. Chemla, T. C. Damen, A. C. Gossard,
W. Wiegmann, T. H. Wood, and C. A. Burrus, Phys. Rev. Lett. {\bf
53}, 2173 (1984).
\bibitem{bernardini97} F. Bernardini, V. Fiorentini, and D.
Vanderbilt, Phys. Rev. B {\bf 56}, R10024 (1997).
\bibitem{Waltereit00} P. Waltereit, O. Brandt, A. Trampert, H. T. Grahn, J. Menniger,
M. Ramsteiner, M. Reiche, and K. H. Ploog, Nature {\bf 406}, 865 (2000).
\bibitem{founta05} S. Founta, F. Rol, E. Bellet-Amalric, J. Bleuse,
B. Daudin, B. Gayral, and H. Mariette, Appl. Phys. Lett. {\bf 86},
171901 (2005).
\bibitem{garro05} N. Garro, A. Cros, J. A. Budagosky, A. Cantarero,
A. Vinattieri, M. Gurioli, S. Founta, H. Mariette, and B. Daudin,
submitted to Appl. Phys. Lett. {\bf 87}, 011101 (2005).
\bibitem{rol07} F. Rol, S. Founta, H. Mariette, B. Daudin, L. S. Dang, J. Bleuse, D. Peyrade,
J.-M. Gérard, y B. Gayral, Phys. Rev. B {\bf 75}, 125306 (2007).
\bibitem{founta07} S. Founta, C. Bougerol, H. Mariette, B. Daudin, y P. Venn\'egu\`es, J. Appl. Phys. {\bf 102}, 074304 (2007).
\bibitem{schulz09} S. Schulz, A. Barube, and E. P. O'Reilly, Phys. Rev. B {\bf 79}, 081401(R) (2009).
\bibitem{marquardt09} O. Marquardt, T. Hickel, abd J. Neugebauer, J. Appl. Phys. {\bf 106}, 083707 (2009).
\bibitem{schuh12} K. Schuh, S. Bartherl. O. Marquardt, T. Hickel, J. Neugebauer, G. Czycholl, and F. Jahnke, Appl. Phys. Lett. {\bf 100}, 092103 (2012).
\bibitem{schulz12} S. Schulz, M. A. Caro, and E. P. O'Reilly, Appl. Phys. Lett. {\bf 101}, 113107 (2012).
\bibitem{corfdir09} P. Corfdir, P. Lefebvre, J. Levrat, A. Dussaigne, J. D. Gani\`ere, D. Martin, J. Riti\'c, T. Zhu, N. Grandjean, and B. Deveaud-Pl\'edran, J. Appl. Phys. {\bf 105}, 043102 (2009).
\bibitem{jacopin11} G. Jacopin, L. Rigutti, L. Largeau, F. Fortuna, F. Furtmayr, F. H. Julien, M. Eickhoff, and M. Tchernycheva, J. Appl. Phys. {\bf 110}, 064313 (2011).
\bibitem{eshelby57} Eshelby, J. D.
Proc. R. S. Series A-Math. Phys. Sc. {\bf 241}, 376-396 (1957).
\bibitem{chuang96} S. L. Chuang y C. Chang, Phys. Rev. B {\bf 54}, 2491-2504 (1996).
\bibitem{yanvoon09} L. C. Lew Yan Voon and M. Willatzen, {\it The k·p method: electronic structure of semiconductors} (Springer, Berlin, 2009).
\bibitem{andreev00} A. D. Andreev and E. P. O'Reilly, Phys. Rev. B
{\bf 62}, 15851 (2000); {\it ibit}, Appl. Phys. Lett. {\bf 79},
521 (2001).
\bibitem{hille14} P. Hille, J. M\"u{\ss}ener, P. Becker, M. de la Mata, N. Rosemann, C. Mag\'en, J. Arbiol, J. Teubert, S. Chatterjee, J. Sch\"ormann, and M. Eickhoff, Appl. Phys. Lett. {\bf 104}, 102104 (2014).
\end{thebibliography}


\end{document}